%% file: main.tex
\documentclass[conference]{IEEEtran}
\IEEEoverridecommandlockouts

\input{preamble}
\input{preamble/preprint}

\DeclareSourcemap{
  \maps[datatype=bibtex, overwrite]{
    \map{
      \perdatasource{bib/sources.bib}
      \step[fieldset=keywords, fieldvalue=source, append]
    }
  }
}

\DeclareSortingScheme{manualsort}{
    \sort[final]{ \field{sortkey} } 
    \sort{\citeorder}
}

\title{A Reference Architecture of\\ Reinforcement Learning Frameworks}

\author{
     \IEEEauthorblockN{Xiaoran Liu\IEEEauthorrefmark{1}\orcidlink{0009-0000-9908-7406},
     Istvan David\IEEEauthorrefmark{1}\IEEEauthorrefmark{2}\orcidlink{0000-0002-4870-8433}}
     \IEEEauthorblockA{\IEEEauthorrefmark{1}McMaster University, Hamilton, Canada}
     \IEEEauthorblockA{\IEEEauthorrefmark{2}McMaster Centre for Software Certification, Hamilton, Canada}
}

\begin{document}

\placetextbox{0.5}{0.99}{\large\colorbox{gray!3}{\textcolor{WildStrawberry}{\textbf{Author pre-print.}}}}%

\placetextbox{0.5}{0.97}{\large\colorbox{gray!3}{\textcolor{WildStrawberry}{Publication accepted for  \hreff{https://conf.researchr.org/home/icsa-2026}{ICSA 2026}.}}}%

\placetextbox{0.5}{0.05}{\colorbox{gray!3}{\textcolor{WildStrawberry}{Author pre-print. Publication accepted for} \hreff{https://conf.researchr.org/home/icsa-2026}{ICSA 2026}.}}%

\maketitle
\IEEEpeerreviewmaketitle

\input{sections/abstract}
\input{sections/intro}
\input{sections/background}
\input{sections/methodology}
\input{sections/ra}
\input{sections/reconstructing-patterns}
\input{sections/evaluation}
\input{sections/discussion}
\input{sections/conclusion}
\input{sections/ack}

\newrefcontext[labelprefix=F,sorting=manualsort]
\printbibliography[keyword=source,title={RL Frameworks},resetnumbers=false]
\newrefcontext
\printbibliography[notkeyword=source,title={References},resetnumbers=true]

\end{document}

%% file: preamble.tex
\input{preamble/packages}
\input{preamble/commands}

\input{preamble/settings}

%% file: preamble/packages.tex
\usepackage[utf8]{inputenc}
\usepackage{ifthen}
\usepackage[hidelinks]{hyperref}

\usepackage[dvipsnames]{xcolor}
\usepackage{soul}
\usepackage[normalem]{ulem}
\usepackage{graphicx}

\usepackage{etoolbox}
\usepackage{environ}

\usepackage{epsf,picinpar}
\usepackage{varioref}
\usepackage{fdsymbol}

\usepackage{enumitem}

\usepackage[backend=biber,defernumbers=true,style=ieee,maxbibnames=99,maxcitenames=2,mincitenames=1,citestyle=numeric-comp,style=ieee]{biblatex}

\usepackage{pifont}

\usepackage{orcidlink}
\usepackage{booktabs}

\usepackage{listings}

\usepackage{multicol}
\usepackage{multirow}
\usepackage{makecell}
\usepackage{caption}
\usepackage{subcaption}
\usepackage[framemethod=TikZ]{mdframed}

\usepackage{dblfloatfix}
\usepackage{nicefrac}

\usepackage[color=red!50!white,textsize=scriptsize]{todonotes}

\usepackage{fontawesome5}

\usepackage{makecell}

\usepackage{url}

\usepackage[nomessages]{fp}
\usepackage[scale=0.9]{inconsolata}
\usepackage{float}

%% file: preamble/commands.tex
\newcommand{\secref}[1]{Sec.~\ref{#1}}
\newcommand{\figref}[1]{Fig.~\ref{#1}}
\newcommand{\tabref}[1]{Tab.~\ref{#1}}

\newcommand{\TODO}[1]{
    \ifthenelse{\boolean{showannotations}}%
    {\ifthenelse{\equal{#1}{}}{\textcolor{VioletRed}{TODO}}{\textcolor{VioletRed}{TODO:~{#1}}}}%
    {}%
}

\newcommand{\assignedto}[1]{%
    \ifthenelse{\boolean{showannotations}}%
    {\textbf{\noindent\ding{46}\textcolor{white}{~\colorbox{\assignementcolor}{Assigned to:}}~\textcolor{\assignementcolor}{#1}\\}%
    }
    {}
}

\newcommand{\component}[1]{%
    \textit{#1}%
}

\newcommand{\componentref}[2]{%
    \hyperref[#1]{\component{#2}}%
}

\newcommand{\componentrefplain}[2]{%
    \hyperref[#1]{#2}%
}

\newcommand{\code}[1]{%
    \texttt{#1}%
}

\newcommand{\rem}[1]{%
    \ifthenelse{\boolean{showannotations}}%
    {\textcolor{\oldtextcolor}{\st{#1}}}%
    {}%
}

\newcommand\add[1]{%
    \ifthenelse{\boolean{showannotations}}%
    {\textcolor{\newtextcolor}{{#1}}}%
    {#1}%
}

\NewEnviron{removeblock}{%
    \ifthenelse{\boolean{showannotations}}%
    {\color{red}\BODY}%
    {}%
}

\NewEnviron{noteblock}{%
    \ifthenelse{\boolean{showannotations}}%
    {\phantom{}\color{gray}%
    
        NOTES:
        
        \BODY
    }%
    {}%
}

\newcommand{\redacted}[1]{%
    \textit{\colorbox{gray!50}{\textless{}Redacted\textgreater{}}}%
}

\newcommand\addblockbegin{%
    \ifthenelse{\boolean{showannotations}}%
    {\color{\newtextcolor}}%
    {}%
}

\newcommand\addblockend{%
    \ifthenelse{\boolean{showannotations}}%
    {\color{black}}%
    {}%
}

\newcommand\rep[2]{%
    \ifthenelse{\boolean{showannotations}}%
    {\rem{#1}~\add{#2}}%
    {#2}%
}

\makeatletter
 \if@todonotes@disabled
 
 \else
 
 \fi
 \makeatother

\makeatletter
 \if@todonotes@disabled
 \newcommand{\donemarker}[2][green]{#2}
 \else
 \newcommand{\donemarker}[2][green]{%
 {\ifthenelse{\equal{#2}{}}{\todo[color={#1}]{DONE UNTIL HERE.}}%
 {\todo[color={#1}]{#2}}}%
 }
 \fi
 \makeatother

\newcommand{\smallparagraph}[1]{
    \noindent\textit{#1.}%
}%

\newcounter{tempOcounter}
\stepcounter{tempOcounter}
\newcounter{tempRcounter}
\stepcounter{tempRcounter}

\newenvironment{recommendationframe}[0]
  {\mdfsetup{
    frametitleaboveskip=-\ht\strutbox,
    frametitlealignment=\center
    }
  \begin{mdframed}[nobreak=true]%
  }
  {\end{mdframed}\stepcounter{tempRcounter}\vspace{-0.75em}}

\newcolumntype{L}[1]{>{\raggedright\let\newline\\\arraybackslash\hspace{0pt}}m{#1}}
\newcolumntype{C}[1]{>{\centering\let\newline\\\arraybackslash\hspace{0pt}}m{#1}}
\newcolumntype{R}[1]{>{\raggedleft\let\newline\\\arraybackslash\hspace{0pt}}m{#1}}

\newcommand{\colheight}{-.25}
\newlength{\maxlen}

\newcommand{\numberofstudies}{18}

\newcommand{\databar}[4][\numberofstudies]{%
    \settowidth{\maxlen}{#3}%
    \addtolength{\maxlen}{\tabcolsep}%
    \ifthenelse{\equal{#1}{}}{%
        \FPeval{perc}{round(100.0 / \studynumberdefault * #2, 1)}%
    }%
    {%
        \FPeval{perc}{round(100.0 / #1 * #2, 1)}%
    }%
    \FPeval\result{round(\perc / #3, 1)}%
    \rlap{\color{#4}\hspace*{-.5\tabcolsep}\rule[\colheight\ht\strutbox]{\result\maxlen}{1.2\ht\strutbox}}%
    \makebox[\dimexpr\maxlen-\tabcolsep][l]{#2 (\perc\%)}%
}

\newcommand\xofy[3]{%
    \FPeval{perc}{round(100.0 / #2 * #1, 1)}%
    \ifthenelse{\equal{#3}{}}{%
        {#1} of {#2} (\perc\%)%
    }%
    {%
        {#1} of {#2} {#3} (\perc\%)%
    }%
}

\newcommand\perc[2]{%
    \FPeval{perc}{round(100.0 / #2 * #1, 1)}%
    \perc%
}

\definecolor{barcolor}{HTML}{85d4ff}
\definecolor{subbarcolor}{HTML}{8fff85}
\definecolor{subsubbarcolor}{HTML}{d8d8d8}

%% file: preamble/settings.tex
\newboolean{showannotations}
\setboolean{showannotations}{true} 

\hypersetup{pdfborder={0 0 0}}

\newcommand{\newtextcolor}{blue}
\newcommand{\oldtextcolor}{red}
\newcommand{\assignementcolor}{orange}
\definecolor{highlightcolor}{rgb}{.99, 1, .0}
\sethlcolor{highlightcolor}
\definecolor{orcidlogocol}{HTML}{A6CE39}

\mdfsetup{%
   backgroundcolor=gray!5,
   middlelinewidth=1pt,
   roundcorner=7pt}



%% file: preamble/preprint.tex
\usepackage{eso-pic}
\definecolor{linkblue}{HTML}{006fbd}
\newcommand{\hreffinternal}[3]{\href{#1}{\textcolor{#3}{#2}}}
\newcommand{\hreff}[2]{\hreffinternal{#1}{#2}{linkblue}}

\newcommand{\placetextbox}[3]{
  \setbox0=\hbox{#3}
  \AddToShipoutPictureFG*{
    \put(\LenToUnit{#1\paperwidth},\LenToUnit{#2\paperheight}){\vtop{{\null}\makebox[0pt][c]{#3}}}%
  }%
}%

%% file: sections/abstract.tex
\begin{abstract}
The surge in reinforcement learning (RL) applications gave rise to diverse supporting technology, such as RL frameworks. However, the architectural patterns of these frameworks are inconsistent across implementations and there exists no reference architecture (RA) to form a common basis of comparison, evaluation, and integration.
To address this gap, we propose an RA of RL frameworks. Through a grounded theory approach, we analyze 18 state-of-the-practice RL frameworks and, by that, we identify recurring architectural components and their relationships, and codify them in an RA. To demonstrate our RA, we reconstruct characteristic RL patterns. Finally, we identify architectural trends, e.g., commonly used components, and outline paths to improving RL frameworks.
\end{abstract}


%% file: sections/intro.tex
\section{Introduction}

Reinforcement learning (RL)~\cite{sutton1998reinforcement} has become one of the most widely used machine learning (ML) techniques~\cite{figueiredoprudencio2024survey}, being adopted in an expanding array of fields from digital twins~\cite{david2023automated} to model-driven engineering~\cite{dagenais2025complex}, outside of its traditional application area of robotics~\cite{kober2013reinforcement}.
In RL, learning is achieved through a process of trial-and-error, in which an agent takes actions and assesses the utility of those actions, to reinforce the beneficial ones. Virtual training frameworks~\cite{kim2021survey} offer a safe and cost-efficient way to train RL agents, by modeling the real world and allowing the agents to interact with this model during training, instead of interacting with the real world~\cite{liu2025ai}.

In response to the surge in RL applications~\cite{terven2025deep,pippas2025evolution}, numerous RL frameworks have appeared~\cite{nouwoumindom2023comparison}.
However, due to the lack of common design guidelines and standards, RL frameworks exhibit diverse architectures and organization of components. This diversity gives rise to inconsistent abstractions among RL frameworks, hinders the reuse of solutions across frameworks, and poses a challenging learning curve for adopters.
Although partial architectural solutions exist, e.g., abstractions for distributed RL~\cite{liang2018rllib}, support for modularizing RL algorithms~\cite{hoffman2020acme}, and decoupling simulators in specific RL frameworks~\cite{schuderer2021sim-env}, the need for a comprehensive architectural understanding of RL remains unaddressed. 
The closest to addressing this need is the work by \textcite{ntentos2024supporting} who propose an Architectural Design Decisions model for developing RL architectures.

To understand the architectures of actual RL frameworks, a reference architecture (RA) of \textit{existing implementations} is needed. Reflecting on the current practices in RL frameworks allows for identifying key architectural tendencies, limitations, and opportunities for improvement.
Therefore, in this work, we analyze 18 frequently used open-source RL frameworks, derive an RA from our analysis, and demonstrate its utility by reconstructing the architectures of characteristic RL patterns.

The importance of comprehensive RAs in RL cannot be overstated. Software engineers and ML developers rely on RL frameworks when they develop and integrate RL functionality into production software systems. Without proper architectural understanding, the quality assessment~\cite{nahar2023meta-summary}, dependency management~\cite{morovati2024common}, certification~\cite{tambon2022how}, and delivery~\cite{baier2019challenges} of these systems becomes a formidable challenge.

\phantom{}

\smallparagraph{Contributions} We make the following contributions.
\begin{itemize}[leftmargin=1em]
    \item We develop a \textbf{reference architecture} for RL environments;
    \item we \textbf{clarify RL architectural concepts} that are often used interchangeably and incorrectly;
    \item we \textbf{reconstruct RL patterns} to demonstrate the RA;
    \item we \textbf{identify architectural tendencies} in RL frameworks.
\end{itemize}

Our contributions benefit (i) RL framework developers by clearly delineating architectural components and providing a blueprint for mapping RL processes onto these components; (ii) adopters by providing a basis for comparing RL frameworks; and (iii) ML engineers by aiding modularity and reusability of their RL pipelines.

\phantom{}\\[-1.75em]

\smallparagraph{Open science}
To enable the independent verification and reuse of our results, we publish a data package as an Open Research Object on Zenodo: \url{https://zenodo.org/records/18637532}.

%% file: sections/background.tex
\section{Background and Related Work}\label{sec:background-relwork}

\subsection{Reinforcement learning and supporting infrastructure}

Reinforcement learning (RL)~\cite{sutton1998reinforcement} is a machine learning (ML) paradigm formalized by Markov decision processes~\cite{puterman1990markov}, in which the agent interacts with the environment to learn the optimal strategies for making sequential decisions.
The agent observes the current state and selects an action according to its decision function---often referred to as a policy---which maps its experienced history of observations to the action.
The environment transitions to a new state and produces a reward based on the agent’s action.
The RL agent refines its decision function through an iterative learning process that requires large volumes of interaction with the environment.

This interaction can be operationalized in physical or virtual environments.
Agents can be trained directly on physical systems, e.g., learning manipulation skills on actual robotic hardware~\cite{kalashnikov2018scalable}, or refining control policies in autonomous vehicles~\cite{kendall2019learning}. However, real-world training is challenged by safety risks, high costs, and privacy constraints~\cite{zhou2017machine}.
Virtual environments overcome these barriers by situating the training process in the virtual space, i.e., in silico.
By that, virtual environments enable safe and cost-efficient agent learning.

\subsection{Terminology blurring in RL infrastructures}\label{sec:blurring}
RL development depends on more than the environment alone. It requires additional components, such as training configurations, learning algorithms, and supporting utilities for data collection, logging, and result evaluation. These broader systems---commonly referred to as frameworks---integrate the environment with the necessary training-related components.

The architectural boundaries that distinguish what constitutes an environment, a framework, or a system are loosely defined in RL practice.
First, the distinction between the environment and the simulator is often blurred.
For example, CARLA is an open-source autonomous driving simulator featuring custom-designed digital assets (e.g., vehicles, buildings, road layouts) that reflect real-world scales and properties~\cite{dosovitskiy2017carla}. However, it is frequently referred to as an environment~\cite{kusmenko2022model-driven, czechowski2025deep}.
Simulators are sometimes quoted as ideal environments for learning~\cite{buhet2019conditional}, demonstrating heavy terminological blurring.
Second, RL algorithms and frameworks are often not properly separated. For example, Dopamine is a research framework for fast prototyping of RL algorithms \cite{castro18dopamine}, yet is referenced as the algorithm libraries in related literature~\cite{kusmenko2022model-driven}.

The lack of clarity creates a barrier to engineering RL frameworks, and underscores the need for an RA that clarifies how architectural primitives are organized into coarser-grained entities, e.g., environment, framework core, framework.

In this paper, unless ambiguous, we use the term \textit{RL framework} inclusive of RL environments and utilities.

\subsection{Related work}
Recent work has begun to address terminological and architectural ambiguities in RL frameworks from different perspectives.
However, there is no general reference architecture that captures common patterns across RL frameworks.

\textcite{schuderer2021sim-env} propose Sim-Env, a workflow and tool for decoupling OpenAI Gym~\cite{brockman2016openai} environments from simulators and models to allow for swapping RL environments while preserving the underlying simulator. 
However, their approach is focused on simulation concerns in a simple single-agent environment, and ignores various other flavors of RL, e.g., multi-agent reinforcement learning (MARL)~\cite{zhang2021multi-agent}.

\textcite{ntentos2024supporting} propose an Architectural Design Decisions (ADDs) model for RL architectures, identifying decision options, relations, and decision drivers for training strategies, such as single versus multi-agent configurations, and checkpoint usage.
Their work provides valuable guidance on architectural choices based on academic and gray literature; our work further investigates the RL architecture by deconstructing existing implementations.

\textcite{balhara2022survey} conduct a systematic literature review (SLR) to analyze different deep reinforcement learning (DRL) algorithms and their architectures. 
However, they focus on DRL algorithm architectures and neural network (NN) structures, rather than proposing a general architecture. 

Some works propose architectures for specific RL toolkits. For example, \textcite{hu2022marllib} present MARLlib's MARL architecture, \textcite{hoffman2020acme} present Acme's distributed learning architecture.
However, these proposals are implementation-specific and do not generalize across different RL frameworks.

To address these limitations, in this work, we propose a general RA for RL by analyzing existing RL frameworks.

%% file: sections/methodology.tex
\section{Methodology}\label{sec:methodology}

In this section, we design a study for recovering architectures of RL frameworks.
We use grounded theory (GT), the method of inductive generation of a theory (here, a general architecture of RL frameworks) from data~\cite{glaser1967discovery}. GT involves simultaneous data collection and analysis through iterative interpretation, aiming to construct a theory rooted in the collected data~\cite{ntentos2024supporting}.
The appeal of GT lies in its general applicability over different types of data, including qualitative, quantitative, semi-structured, interviews, etc.~\cite{glaser1998doing}; and in this work, specifically, source code and design documentation encountered in open-source repositories.

GT has been proven to be of high utility in recovering architectures~\cite{tamburri2018general} and architectural decision points~\cite{ntentos2024supporting} previously.

\subsection{Iterative coding and data collection}

We employ the Strauss-Corbin flavor of GT~\cite{corbin2015basics}, in which three coding phases of open, axial, and selective coding are used, to produce a detailed, explanatory theory of RL frameworks' architectures.
In the open coding phase, we review source code, configuration files, and documentation of RL frameworks, categorizing implementation details into conceptual labels, such as \textit{Algorithm}, \textit{Optimizer}, and \textit{Learner}.
In the axial coding phase, we cluster related labels into architectural components and identify component interactions.
For example, we group \textit{Algorithm}, \textit{Optimizer}, and \textit{Learner} under the \componentref{sec:ra-learner}{Learner} component due to their strongly related functionality, and identify their relationships with the \componentref{sec:ra-buffer}{Buffer} and \componentref{sec:ra-function-approximator}{Function Approximator} components to form the \componentref{sec:ra-agent}{Agent} component.
In the selective coding phase, we refine and integrate components into the theory of RL frameworks' architecture, encompassing environment design, simulator integration, agent-environment interaction, and training orchestration. 

Following the principle of immediate and continuous data analysis of GT~{\cite{stol2016grounded}}, we conduct these coding steps iteratively, i.e., after a coding phase, we constantly compare data, memos, labels, and categories across sources. We implement memos through detailed notes maintained in the analysis spreadsheet, capturing our thought processes, interpretations, and reasoning to ensure traceability of labels and categories to their origins.
To mitigate threats to validity stemming from the researchers' biases, we facilitate constant cross-verification steps among the researchers, and discussions when agreement is not immediate.

\subsection{Sampling}
We sample RL environments and frameworks.
At this stage of the study, we could only rely on the usual imprecise classification of RL systems (see \secref{sec:blurring}) and sampled both \textit{environments} and \textit{frameworks}, and reach saturation in both classes of systems---i.e., the point where new data no longer yields new insights to the theory~\cite{glaser1978theoretical}. Saturation in environments is reached after analyzing five of them (\cite{towers2025gymnasium, terry2021pettingzoo, juliani2020, makoviychuk2021isaac, mittal2025isaac}); subsequent environments (\cite{tunyasuvunakool2020, beattie2016deepmind, bellemare2013arcade, bonnet2024jumanji}) confirm the same categories and relationships without adding new architectural elements. Saturation in frameworks is reached after analyzing six of them (\cite{stable-baselines3, rl-zoo3, liang2018rllib, hoffman2020acme, hu2022marllib, bettini2024benchmarl}); subsequent training sources (\cite{dekock2023mava, castro18dopamine, tianshou}) confirm existing categories without yielding new insights.

We drive our sampling through our domain understanding as a heuristic and aim to cover a wide range of intents and usage patterns early on.
We start with Gymnasium~\cite{towers2025gymnasium}, a single-agent RL environment. Subsequently, we open up our analysis to multi-agent environments by sampling PettingZoo~\cite{terry2021pettingzoo}. This allows us to investigate inter-agent coordination mechanisms. Then, we open up our investigation to frameworks beyond RL environments and investigate more complex RL frameworks, such as RLLib~\cite{liang2018rllib}. This allows us to investigate how multi-agent coordination extends beyond training, specifically into policy management and execution.

Eventually, we sample and analyze 18 RL frameworks. This sample features the most widely used RL systems in both research and practice, evidenced by a mix of peer-reviewed literature~\cite{kusmenko2022model-driven,kim2021survey,liu2024acceleration} and community-curated collections.\footnote{\url{https://github.com/awesomelistsio/awesome-reinforcement-learning}}

%% file: sections/ra.tex
\section{Reference Architecture of RL Frameworks}\label{sec:ra}

We now present the reference architecture (RA) of RL frameworks we developed through our empirical inquiry.

The high-level overview of the RA is shown in \figref{fig:high-level}. It contains six top-level components organized into four component groups: the \componentref{sec:ra-framework}{Framework} (\secref{sec:ra-framework}), \componentref{sec:ra-framework-core}{Framework Core} (\secref{sec:ra-framework-core}), \componentref{sec:ra-environment}{Environment} (\secref{sec:ra-environment}), and \componentref{sec:ra-utilities}{Utilities} (\secref{sec:ra-utilities}).
In the following, we elaborate on each of these components.
For each component, we provide a detailed internal architectural overview, and report which components can be found in various RL frameworks as separate entities. Often, components are implemented under different names, or in an aggregation or amalgamation with other components.
Such details are provided in the replication package.

The analyzed RL systems often substantially differ in scope. Some provide only a \componentref{sec:ra-framework-core}{Framework Core}---these are typically the ones that are colloquially referred to as ``environments,'' e.g., Gymnasium~\cite{towers2025gymnasium}. Others implement additional services that typically fall into the \componentref{sec:ra-utilities}{Utilities} (\secref{sec:ra-utilities}) category of components and delegate the responsibility of defining agents and environments to the former class of RL systems---these are typically the ones that are colloquially referred to as ``frameworks,'' e.g., Stable Baselines3~\cite{stable-baselines3}.
As RL is formally defined as a Markov decision process of an \componentref{sec:ra-agent}{Agent} in an \componentref{sec:ra-environment}{Environment}~\cite{sutton1998reinforcement}, it is these two components that are truly essential in RL systems, and one can encounter simplistic RL experiments with only these two components. Nonetheless, reasonably complex RL experiments necessitate additional services, e.g., visualization in 3D simulated environments, or data persistence for long-running experiments.

\input{sections/ra/system}
\input{sections/ra/framework}
\input{sections/ra/environment}
\input{sections/ra/utils}

%% file: sections/ra/system.tex
\begin{figure}[t]
    \centering
    \includegraphics[width=0.9\linewidth]{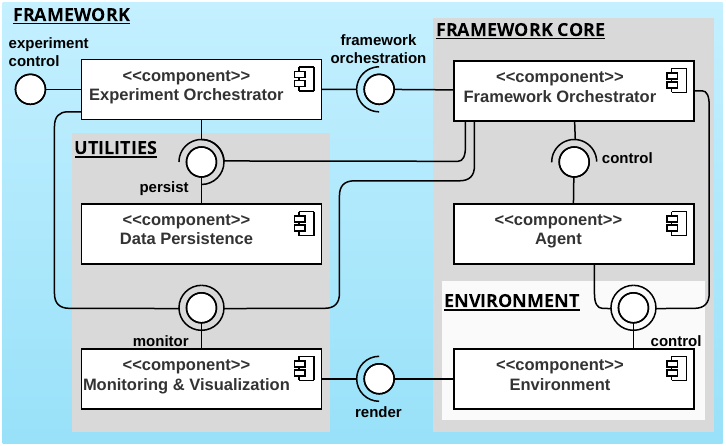}
    \caption{High-level architectural view of RL frameworks. (Component groups are \ul{underlined} and correspond to the subsections of \secref{sec:ra}.)}
    \label{fig:high-level}
\end{figure}

\subsection{Framework}\label{sec:ra-framework}

The \component{Framework} comprises the user-facing \componentref{sec:ra-experiment-orchestrator}{Experiment Orchestrator} component, the \componentref{sec:ra-framework-core}{Framework Core}, and the \componentref{sec:ra-utilities}{Utilities}. Its primary responsibility is to enable users to configure and execute experiments. An experiment is a collection of training or evaluation executions, with particular hyperparameters and configurations~\cite{zaharia2018accelerating}.
The execution of individual runs within an experiment is handled by the \componentref{sec:ra-framework-core}{Framework Core}.

\begin{figure}[h]
    \centering
    \includegraphics[width=\linewidth]{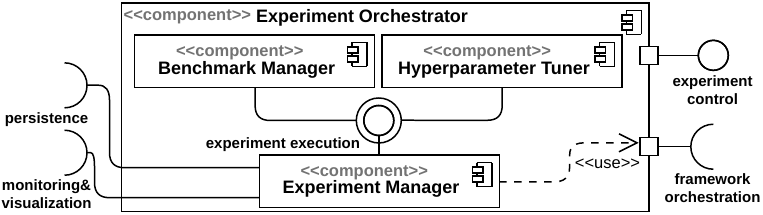}
    \caption{Experiment Orchestrator}
    \label{fig:ra-experiment-orchestrator}
\end{figure}

\input{tables/experiment-orchestrator}

\subsubsection{Experiment Orchestrator}\label{sec:ra-experiment-orchestrator}
Provides the primary interface to users for defining and running experiments, e.g., training agents, tuning hyperparameters, or benchmarking algorithms. It orchestrates the high-level experiment process with the optional hyperparameter tuning and benchmarking steps before the sequences of experiments are executed.
The \component{Experiment Orchestrator} consists of three components (\figref{fig:ra-experiment-orchestrator}, \tabref{tab:ra-experiment-orchestrator}).

\paragraph{Experiment Manager}\label{sec:ra-experiment-manager}
The \component{Experiment Manager} sets up the experiment for execution, prepares the \component{Data Persistence}, \component{Monitor \& Visualization} components, and delegates the experiment execution to the \componentref{sec:ra-framework-orchestrator}{Framework Orchestrator}.
It relies on two \componentref{sec:ra-utilities}{Utilities} components to save and load experiment state (via the \componentref{sec:ra-data-persistence}{Data Persistence} component), and to monitor and visualize the results (via the \componentref{sec:ra-monitoring-and-visualization}{Monitoring \& Visualization} component).

\paragraph{Hyperparameter Tuner}\label{sec:ra-hyperparameter-tuner}
Automates the search for the optimal hyperparameter configurations by sampling the hyperparameter space.
It generates candidate hyperparameters (e.g., via grid search~\cite{larochelle2007empirical}, random search~\cite{bergstra2012random}, or Bayesian optimization~\cite{snoek2012practical}), and passes them to the \componentref{sec:ra-experiment-manager}{Experiment Manager} to execute experiments for evaluation.
Upon executing the experiment, the \component{Hyperparameter Tuner} analyzes the results to guide subsequent sampling iterations.
This process continues until stopping criteria are met, e.g., convergence to user-defined objectives or exhaustion of the search budget.

The implementation is often provided by specialized third-party libraries.
For example, RL-Zoo3~\cite{rl-zoo3} uses Optuna~\cite{akiba2019optuna}, and RLlib~\cite{liang2018rllib} uses Ray Tune~\cite{liaw2018tune}.

\paragraph{Benchmark Manager}\label{sec:ra-benchmark-manager}
Enables the evaluation of different RL algorithms under consistent experimental settings.
Executes experiments (via the \componentref{sec:ra-experiment-manager}{Experiment Manager}) that share a base configuration but vary in algorithms or policies.
An example of this component is given by the \code{Benchmark} module in the BenchMARL framework~\cite{bettini2024benchmarl}.\footnote{Detailed pointers to modules and source code in the analyzed frameworks are available in the data package.}

%% file: tables/experiment-orchestrator.tex
\begin{table}[h]
\centering
\caption{Experiment Orchestrator Components}
\label{tab:ra-experiment-orchestrator}
\renewcommand{\arraystretch}{0.9}
{\scriptsize
\begin{tabular}{@{}p{2.6cm}p{5.9cm}@{}}
\toprule
\multicolumn{1}{c}{\textbf{Component}} & \multicolumn{1}{c}{\textbf{Frameworks}} \\ \midrule

Benchmark Manager & \cite{bettini2024benchmarl}\\

Experiment Manager & \cite{juliani2020} \cite{stable-baselines3} \cite{rl-zoo3} \cite{liang2018rllib} \cite{hoffman2020acme} \cite{hu2022marllib} \cite{bettini2024benchmarl} \cite{dekock2023mava} \cite{castro18dopamine} \cite{tianshou}\\

Hyperparameter Tuner & \cite{rl-zoo3} \cite{liang2018rllib} \cite{hu2022marllib}\\

\bottomrule
\end{tabular}}
\end{table}

%% file: sections/ra/framework.tex
\subsection{Framework Core}\label{sec:ra-framework-core}

The \component{Framework Core} coordinates the learning process. It encapsulates the \componentref{sec:ra-environment}{Environment} component group, as well as the \componentref{sec:ra-framework-orchestrator}{Framework Orchestrator}, and the \componentref{sec:ra-agent}{Agent} components.

\subsubsection{Framework Orchestrator}\label{sec:ra-framework-orchestrator}
\begin{figure}[t]
    \centering
    \includegraphics[width=0.85\linewidth]{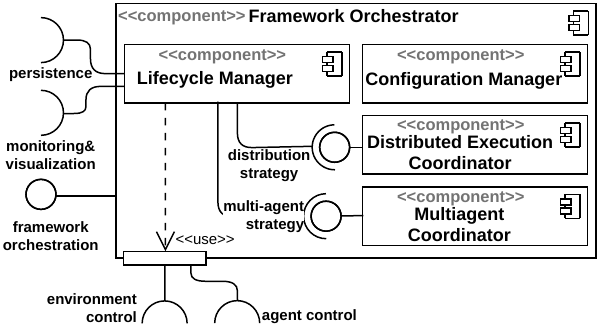}
    \caption{Framework Orchestrator}
    \label{fig:ra-framework-orchestrator}
\end{figure}

Receives experiment requests from the \componentref{sec:ra-experiment-manager}{Experiment Manager} and orchestrates the framework to execute the training or evaluation runs.
It loads configurations, controls the training and evaluation lifecycle, allocates resources for distributed execution if needed, and coordinates multi-agent execution when applicable.
The framework orchestrator consists of four components (\figref{fig:ra-framework-orchestrator}, \tabref{tab:ra-framework-orchestrator}).

\paragraph{Lifecycle Manager} \label{sec:ra-lifecycle-manager}
Controls the execution of the training and evaluation loops.
It initializes required components (via the \componentref{sec:ra-configuration-manager}{Configuration Manager}), handles episode termination and truncation from the environment, monitors global stopping criteria, e.g., maximum timesteps or convergence thresholds, and triggers lifecycle events. Notably, it coordinates the interaction between the \componentref{sec:ra-agent}{Agent} and the \componentref{sec:ra-environment}{Environment}.
It saves and loads the training state via the \componentref{sec:ra-data-persistence}{Data Persistence} component, and tracks per-step metrics and activates rendering via the \componentref{sec:ra-monitoring-and-visualization}{Monitoring \& Visualization}.

The \component{Lifecycle Manager} has two patterns to control the agent-environment interaction. In the first pattern, the \component{Lifecycle Manager} queries actions from the \componentref{sec:ra-agent}{Agent}, forwards them to the \componentref{sec:ra-environment}{Environment}, and sends resulting data back to the \componentref{sec:ra-agent}{Agent} (e.g., Acme~\cite{hoffman2020acme}).
Alternatively, the \component{Lifecycle Manager} actuates the \componentref{sec:ra-agent}{Agent} to interact with \componentref{sec:ra-environment}{Environment} and schedules the execution of actions and learning updates (e.g., RLlib~\cite{liang2018rllib}).

\input{tables/framework-orchestrator}

Implementations of the \component{Lifecycle Manager} differ across
frameworks. Some implement the manager in a single module, while others distribute lifecycle logic into multiple modules.
For example, Acme~\cite{hoffman2020acme} uses the \code{EnvironmentLoop} module to coordinate the interaction between the \componentref{sec:ra-environment}{Environment} and the \componentref{sec:ra-agent}{Agent}.
Isaac Lab~\cite{mittal2025isaac} implements lifecycle management through two managers. The \code{EventManager} orchestrates operations based on different simulation events across the lifecycle; and the \code{TerminationManager} computes termination signals.

\paragraph{Configuration Manager} \label{sec:ra-configuration-manager}
Loads and validates the configurations for the learning process from various sources (e.g., YAML files, JSON files, command-line arguments). Configurations specify which algorithm to use, which environment to instantiate, the execution mode (e.g., distributed vs. non-distributed), and resource allocation preferences.

About half of the sampled RL frameworks use custom configuration managers, and half integrate third-party libraries. Hydra~\cite{yadan2019hydra} is the most common third-party library, adopted by, e.g., Isaac Lab~\cite{mittal2025isaac}, BenchMARL~\cite{bettini2024benchmarl}, and Mava~\cite{dekock2023mava}.

\paragraph{Multi-Agent Coordinator}\label{sec:ra-multi-agent-coordinator} Manages interactions among agents in multi-agent RL, and coordinates them, including the management of policy assignment, i.e., determining which policy controls each agent and whether policies are shared across agents. It also constructs joint actions from individual agent policies, and coordinates agent-to-agent communication when required.

\paragraph{Distributed Execution Coordinator}\label{sec:ra-distributed-execution-coordinator}
Allocates and deploys components across multiple processes, devices, or machines when distributed execution is configured.
It determines the distributed topology by mapping logical components, typically of the \componentref{sec:ra-agent}{Agent} component group (e.g., function approximators, learners, buffers) onto physical resources (e.g., CPUs and GPUs across different machines), deploys these components to their assigned resources, and maintains the metadata (e.g., IP addresses) needed for inter-component communication.

The implementation is typically provided by third-party libraries, e.g., Ray Core~\cite{moritz2018ray} (used by RLlib~\cite{liang2018rllib} and MARLlib~\cite{hu2022marllib}) and Launchpad~\cite{yang2021launchpad} (used by Acme~\cite{hoffman2020acme}).

\subsubsection{Agent}\label{sec:ra-agent}
\begin{figure}[t]
    \centering
    \includegraphics[width=0.8\linewidth]{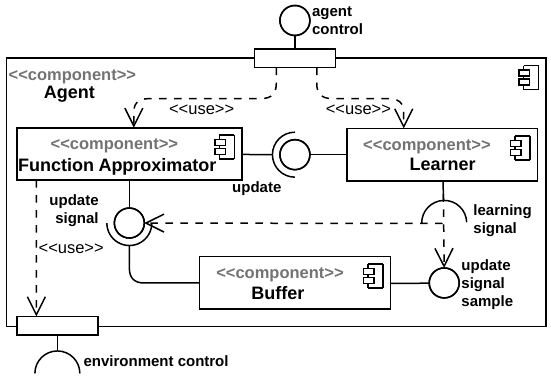}
    \caption{Agent}
    \label{fig:ra-agent}
\end{figure}

The \component{Agent} implements the RL algorithm. It interacts with the \componentref{sec:ra-environment}{Environment} to learn.
The schedule of the learning cycle is controlled by the \componentref{sec:ra-lifecycle-manager}{Lifecycle Manager}, which determines, e.g., when actions are selected, when experience is collected, and when learning updates occur.
Some frameworks provide only the \component{Agent} component and delegate the definition of an \componentref{sec:ra-environment}{Environment} to external libraries through standardized interfaces. For example, Acme provides out-of-the-box integration with the DeepMind Control Suite~\cite{tunyasuvunakool2020} and DeepMind Lab~\cite{beattie2016deepmind}; other environments must implement the \code{dm\_env} interface~\cite{muldal2019dm} to integrate with Acme.
The \component{Agent} consists of three components, as shown in \figref{fig:ra-agent} and \tabref{tab:ra-agent}.

\input{tables/agent}

\paragraph{Function Approximator}\label{sec:ra-function-approximator}
Encodes the agent's decision-making mechanism, which maps states to action selections or value estimates. It selects actions during the \componentref{sec:ra-agent}{Agent}'s interactions with the \componentref{sec:ra-environment}{Environment}, and produces update signals, e.g., experience data in actor-type approximators, or advantage information in critic-type approximators.
The details of the \component{Function Approximator} vary by algorithm type. 
Policy-based methods (e.g., proximal policy optimization (PPO)~\cite{schulman2017proximal}) implement the \component{Function Approximator} as an explicit policy, with a probability-based sampling serving for action selection.
Value-based methods (e.g., Q-learning~\cite{watkins1992q-learning}) implement the \component{Function Approximator} as the value function that estimates expected return for each action in a given state, from which actions are derived (e.g., via $\epsilon$-greedy strategy).
Complex agent architectures may use multiple instances of the component. For example, actor–critic methods~\cite{haarnoja2018soft} instantiate the \component{Function Approximator} twice: as a policy-based actor and as a value-based critic.

The internal representation depends on the data type and state space. For small, discrete spaces, tabular representations are often feasible (e.g., Q-table~\cite{hirashima1999q-learning}). Deep RL frameworks necessitate using a neural network~\cite{sutton1999policy}.

\paragraph{Learner}\label{sec:ra-learner}
Updates the \componentref{sec:ra-function-approximator}{Function Approximator} from learning signals.
It receives these signals either directly from the \componentref{sec:ra-function-approximator}{Function Approximator} (e.g., in the SARSA algorithm~\cite{rummery1994on-line}) or samples it from the \componentref{sec:ra-buffer}{Buffer} (e.g., in the soft actor-critic (SAC) method~\cite{haarnoja2018soft}). Subsequently, it computes algorithm-specific parameter updates (e.g., temporal-difference (TD) errors defined by Bellman equations~\cite{bellman1966dynamic}, or policy gradients~\cite{sutton1999policy}), and applies them to the \componentref{sec:ra-function-approximator}{Function Approximator}.

\paragraph{Buffer}\label{sec:ra-buffer}
Stores update signals and provides the \componentref{sec:ra-learner}{Learner} with sampled batches.
There are two common \component{Buffer} types. A rollout buffer---often used in on-policy RL, e.g., PPO~\cite{schulman2017proximal}---stores short trajectories that are consumed immediately and then discarded. A replay buffer---often used in off-policy RL, e.g., SAC~\cite{haarnoja2018soft}---stores a large pool of transitions and allows the \componentref{sec:ra-learner}{Learner} to request samples during training.

Most RL frameworks implement buffers directly rather than depending on external libraries. Some of the exceptions include Acme~\cite{hoffman2020acme}, which integrates Reverb~\cite{cassirer2021reverb} for distributed settings; and Mava~\cite{dekock2023mava}, which uses Flashbax~\cite{toledo2023flashbax}.

%% file: tables/framework-orchestrator.tex
\begin{table}[t]
\centering
\caption{Framework Orchestrator Components}
\label{tab:ra-framework-orchestrator}
\renewcommand{\arraystretch}{0.9}
{\scriptsize
\begin{tabular}{@{}p{2.6cm}p{5.9cm}@{}}
\toprule
\multicolumn{1}{c}{\textbf{Component}} & \multicolumn{1}{c}{\textbf{Frameworks}} \\ \midrule
Configuration Manager & \cite{mittal2025isaac} \cite{liang2018rllib}  \cite{hu2022marllib} \cite{bettini2024benchmarl} \cite{dekock2023mava} \cite{castro18dopamine} \cite{tianshou}\\

Distributed Exec. Coord. & \cite{liang2018rllib} \cite{hoffman2020acme} \cite{hu2022marllib} \cite{dekock2023mava}\\

Lifecycle Manager & \cite{juliani2020} \cite{stable-baselines3} \cite{rl-zoo3} \cite{liang2018rllib} \cite{hoffman2020acme} \cite{hu2022marllib} \cite{bettini2024benchmarl} \cite{dekock2023mava} \cite{castro18dopamine} \cite{tianshou}\\

Multi-Agent Coord. & \cite{terry2021pettingzoo} \cite{juliani2020} \cite{mittal2025isaac} \cite{liang2018rllib} \cite{hu2022marllib} \cite{bettini2024benchmarl} \cite{tianshou}\\

\bottomrule
\end{tabular}}
\end{table}

%% file: tables/agent.tex
\begin{table}[t]
\centering
\caption{Agent Components}
\label{tab:ra-agent}
\renewcommand{\arraystretch}{0.9}
{\scriptsize
\begin{tabular}{@{}p{1.8cm}p{6.7cm}@{}}
\toprule
\multicolumn{1}{c}{\textbf{Component}} & \multicolumn{1}{c}{\textbf{Frameworks}} \\ \midrule

Buffer & \cite{juliani2020} \cite{stable-baselines3} \cite{rl-zoo3} \cite{liang2018rllib} \cite{hoffman2020acme} 
\cite{hu2022marllib} \cite{bettini2024benchmarl} \cite{dekock2023mava} \cite{castro18dopamine} \cite{tianshou} \\

Func. Approx. & \cite{juliani2020} \cite{stable-baselines3} \cite{rl-zoo3} \cite{liang2018rllib} \cite{hoffman2020acme} 
\cite{hu2022marllib} \cite{bettini2024benchmarl} \cite{dekock2023mava} \cite{castro18dopamine} \cite{tianshou} \\

Learner & \cite{juliani2020} \cite{stable-baselines3} \cite{rl-zoo3} \cite{liang2018rllib} \cite{hoffman2020acme} 
\cite{hu2022marllib} \cite{bettini2024benchmarl} \cite{dekock2023mava} \cite{castro18dopamine} \cite{tianshou}\\

\bottomrule
\end{tabular}}
\end{table}


%% file: sections/ra/environment.tex
\begin{figure}[t]
    \centering
    \includegraphics[width=\linewidth]{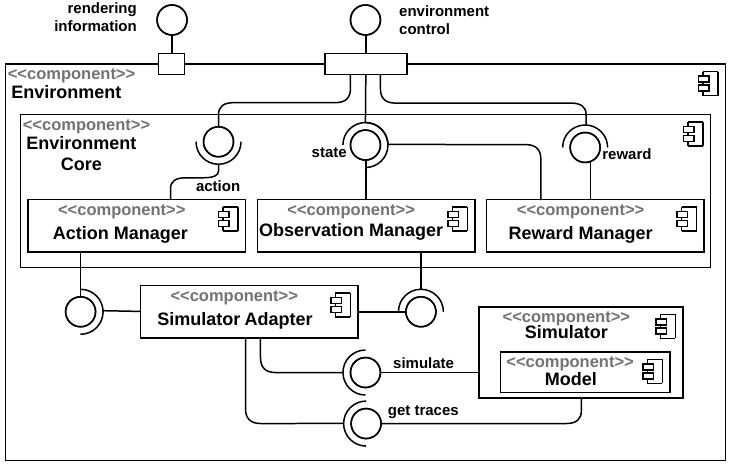}
    \caption{Environment}
    \label{fig:ra-environment}
\end{figure}

\input{tables/environment}

\subsection{Environment}\label{sec:ra-environment}

The \component{Environment} is a group of components with which \componentref{sec:ra-agent}{Agent} interacts.
It chiefly encapsulates the simulation infrastructure that provides a virtual world for interactions.
The \component{Environment} consists of three components (\figref{fig:ra-environment}, \tabref{tab:ra-environment}).

\subsubsection{Environment Core}\label{sec:ra-environment-core}
Exposes the control interface to the \componentref{sec:ra-framework-orchestrator}{Framework Orchestrator}.
It initializes and resets the environment, receives agent actions, updates the environment state, computes observations and rewards, and produces rendered frames for visualization. The \component{Environment Core} may be implemented as a vectorized environment, i.e., multiple parallel environment instances, to improve data collection efficiency~\cite{towers2025gymnasium}.
The \component{Environment Core} includes an \component{Action Manager} to process and apply actions, an \component{Observation Manager} to collect observations from the simulator infrastructure, and a \component{Reward Manager} to compute reward signals~\cite{mittal2025isaac}.

\subsubsection{Simulator}\label{sec:ra-simulator}
The program that encodes the probabilistic mechanism that represents the real phenomenon and executes this probabilistic mechanism over a sufficiently long period of time to produce simulation traces characterizing the actual system~\cite{ross2022simulation}.
At the core of the \component{Simulator}, the physical asset is represented by a \component{Model}. This \component{Model} captures the essential properties of the simulated asset in appropriate detail to consider the results of the simulation representative~\cite{zeigler2018theory}.

RL environments rely on a range of simulators.
For example, Gymnasium~\cite{towers2025gymnasium} supports Box2D~\cite{box2d-github} for 2D physics, Stella~\cite{stella-github} for Atari games, and MuJoCo~\cite{todorov2012mujoco} for control tasks. DeepMind Lab~\cite{beattie2016deepmind} uses the ioquake3 3D engine~\cite{ioquake3-web}.

\subsubsection{Simulator Adapter}\label{sec:ra-simulator-adapter}
Connects the \componentref{sec:ra-environment-core}{Environment Core} with the underlying \componentref{sec:ra-simulator}{Simulator}.
It translates the \componentref{sec:ra-agent}{Agent}'s action to simulation steps, and the \componentref{sec:ra-lifecycle-manager}{Lifecycle Manager}'s instructions for resetting and pausing the simulation. After executing the simulation steps that correspond to the \componentref{sec:ra-agent}{Agent}'s action, the \component{Simulator Adapter} translates the simulation traces to observations, subsequently available for the \componentref{sec:ra-environment-core}{Observation Manager}.

%% file: tables/environment.tex
\begin{table}[t]
\centering
\caption{Environment Components}
\label{tab:ra-environment}
\renewcommand{\arraystretch}{0.9}
{\scriptsize
\begin{tabular}{@{}p{2.1cm}p{6.4cm}@{}}
\toprule
\multicolumn{1}{c}{\textbf{Component}} & \multicolumn{1}{c}{\textbf{Frameworks}} \\ \midrule
Environment Core & \cite{towers2025gymnasium} \cite{terry2021pettingzoo} \cite{juliani2020} \cite{makoviychuk2021isaac} \cite{mittal2025isaac} \cite{tunyasuvunakool2020} \cite{beattie2016deepmind} \cite{bellemare2013arcade} \cite{bonnet2024jumanji} 
\\
Simulator & \cite{towers2025gymnasium} \cite{terry2021pettingzoo} \cite{juliani2020} \cite{makoviychuk2021isaac} \cite{mittal2025isaac} \cite{tunyasuvunakool2020} \cite{beattie2016deepmind} \cite{bellemare2013arcade} \cite{bonnet2024jumanji} 
\\
Simulator Adapter & \cite{towers2025gymnasium} \cite{terry2021pettingzoo} \cite{juliani2020} \cite{makoviychuk2021isaac} \cite{mittal2025isaac} \cite{tunyasuvunakool2020} \cite{beattie2016deepmind} \cite{bellemare2013arcade} 
\\
\bottomrule
\end{tabular}}
\end{table}

%% file: sections/ra/utils.tex
\begin{figure}[t]
    \centering
    \includegraphics[width=0.85\linewidth]{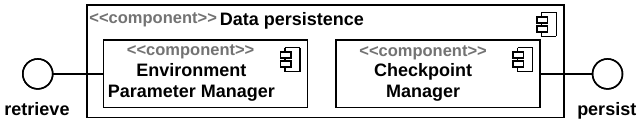}
    \caption{Data Persistence}
    \label{fig:ra-data-persistence}
\end{figure}

\input{tables/data-persistence}

\subsection{Utilities}\label{sec:ra-utilities}
The \component{Utilities} provide services to the rest of the components. 
It has two components: \componentref{sec:ra-data-persistence}{Data Persistence} for state management and experiment resumption, and \componentref{sec:ra-monitoring-and-visualization}{Monitoring \& Visualization} for tracking training progress and visualizing information.

\subsubsection{Data Persistence}\label{sec:ra-data-persistence}
Manages the storage and retrieval of the experiment state. It has two components (\figref{fig:ra-data-persistence}, \tabref{tab:ra-data-persistence}).

\paragraph{Checkpoint Manager}
\label{sec:ra-checkpoint-manager}
Saves and restores the experiment state.
It stores the algorithms' parameters (e.g., policy weights, replay buffer contents), the environment state, and metadata needed for experiment resumption.
The \componentref{sec:ra-lifecycle-manager}{Lifecycle Manager} invokes the checkpointing logic at pre-configured intervals (e.g., every N steps); and the \componentref{sec:ra-experiment-manager}{Experiment Manager} uses it to resume interrupted experiments.

Implementation may be native or provided by a third-party library.
For example, Acme~\cite{hoffman2020acme} implements the \component{Checkpoint Manager} in the \code{Checkpointer} module.
Mava~\cite{dekock2023mava} uses an external checkpoint management library,  Orbax~\cite{orbax-github}.

\paragraph{Environment Parameter Manager}
\label{sec:ra-environment-parameter-manager}
Handles environment parameters that change over time, e.g., difficulty levels in curriculum learning~\cite{bengio2009curriculum} and parameters in domain randomization methods~\cite{tobin2017domain}.
It exposes a retrieval interface that allows the \componentref{sec:ra-lifecycle-manager}{Lifecycle Manager} to query parameters based on the current learning progress. The \componentref{sec:ra-lifecycle-manager}{Lifecycle Manager} uses these parameters when initializing or resetting the environment (e.g., in new training episodes), and for method-specific tasks, e.g., randomizing parameters in the environment~\cite{marcinandrychowicz2020learning}.
For example, ML-Agents~\cite{juliani2020} implements the component in the \code{EnvironmentParameters} module.

\subsubsection{Monitoring \& Visualization}\label{sec:ra-monitoring-and-visualization}
\begin{figure}[t]
    \centering
    \includegraphics[width=0.8\linewidth]{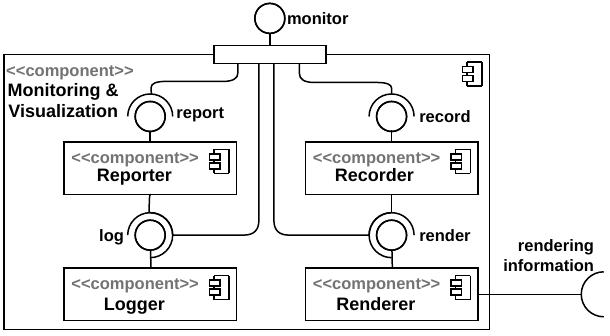}
    \caption{Monitoring and Visualization}
    \label{fig:ra-monitoring-and-visualization}
\end{figure}

Tracks training metrics, generates diagnostic outputs or result summaries, and produces visual representations of agent behavior. 
The frameworks in our sample organize these responsibilities into one larger component. The \component{Logger} produces raw data that other components strongly depend on to track and visualize the training progress.
It consists of four components (\figref{fig:ra-monitoring-and-visualization}, \tabref{tab:ra-monitoring-and-visualization}).

\input{tables/monitoring-and-visualization}

\paragraph{Renderer}\label{sec:ra-renderer}
Generates visual frames of the environment based on data obtained from the \componentref{sec:ra-environment}{Environment}'s \code{rendering information} interface (\figref{fig:ra-environment}).
Common modes include \code{rgb\_array} to return image arrays for programmatic use, and \code{human} to display interactive viewports for humans~\cite{juliani2020}.

\paragraph{Recorder}\label{sec:ra-recorder}
Captures frames via the \componentref{sec:ra-renderer}{Renderer} and assembles them into videos or image sequences.

For example, Gymnasium implements this component by wrapping environments with the \code{RecordVideo} wrapper to capture episodic videos during interaction~\cite{towers2025gymnasium}. Isaac Lab's \code{RecorderManager} records per-step and per-episode frames and exports them through dataset file handlers~\cite{mittal2025isaac}.

\input{tables/component-responsibilities}

\paragraph{Logger}\label{sec:ra-logger}
Records raw experiment data, such as episode returns, losses, and diagnostic information, and persists them for later consumption by the \componentref{sec:ra-reporter}{Reporter}.
Most RL frameworks implement a custom \component{Logger}, but some integrate external experiment tracking tools. For example, Isaac Lab~\cite{mittal2025isaac} and Mava~\cite{dekock2023mava} use Neptune Logger to log metrics, model parameters, and gradients during execution.

\paragraph{Reporter}\label{sec:ra-reporter}
Transforms logged data into human-readable outputs. It is typically used for generating training summaries, performance tables, and learning curves. 

%% file: tables/data-persistence.tex
\begin{table}[t]
\centering
\caption{Data Persistence Components}
\label{tab:ra-data-persistence}
\renewcommand{\arraystretch}{0.9}
{\scriptsize
\begin{tabular}{@{}p{2cm}p{6.5cm}@{}}
\toprule
\multicolumn{1}{c}{\textbf{Component}} & \multicolumn{1}{c}{\textbf{Frameworks}} \\ \midrule
Checkpoint Mgr & \cite{juliani2020} \cite{bonnet2024jumanji} \cite{stable-baselines3} \cite{rl-zoo3} \cite{liang2018rllib} \cite{hoffman2020acme} 
\cite{hu2022marllib} \cite{bettini2024benchmarl} \cite{dekock2023mava} \cite{castro18dopamine} \cite{tianshou}
\\
Env. Parameter Mgr & \cite{juliani2020} \cite{makoviychuk2021isaac} \cite{mittal2025isaac} \cite{bellemare2013arcade}
\\
\bottomrule
\end{tabular}}
\end{table}

%% file: tables/monitoring-and-visualization.tex
\begin{table}[t]
\centering
\caption{Monitoring \& Visualization Components}
\label{tab:ra-monitoring-and-visualization}
\renewcommand{\arraystretch}{0.9}
{\scriptsize
\begin{tabular}{@{}p{0.8cm}p{7.7cm}@{}}
\toprule
\multicolumn{1}{c}{\textbf{Comp.}} & \multicolumn{1}{c}{\textbf{Frameworks}} \\ \midrule

Logger & \cite{towers2025gymnasium} \cite{terry2021pettingzoo} \cite{juliani2020} \cite{makoviychuk2021isaac} \cite{mittal2025isaac} \cite{tunyasuvunakool2020} 
\cite{beattie2016deepmind} \cite{bellemare2013arcade} \cite{bonnet2024jumanji} \cite{stable-baselines3} \cite{rl-zoo3} \cite{liang2018rllib} \cite{hoffman2020acme} \cite{hu2022marllib}
\cite{bettini2024benchmarl} \cite{dekock2023mava} \cite{castro18dopamine} \cite{tianshou}  \\

Recorder & \cite{towers2025gymnasium} \cite{terry2021pettingzoo} \cite{makoviychuk2021isaac} \cite{mittal2025isaac} \cite{beattie2016deepmind} \cite{bellemare2013arcade} \cite{stable-baselines3} \cite{rl-zoo3} \cite{hoffman2020acme} \cite{bettini2024benchmarl} \cite{castro18dopamine}\\

Renderer & \cite{towers2025gymnasium} \cite{terry2021pettingzoo} \cite{juliani2020} \cite{makoviychuk2021isaac} \cite{mittal2025isaac} \cite{tunyasuvunakool2020} \cite{beattie2016deepmind} \cite{bellemare2013arcade} \cite{bonnet2024jumanji}\\

Reporter & \cite{towers2025gymnasium} \cite{terry2021pettingzoo} \cite{juliani2020} \cite{makoviychuk2021isaac} \cite{stable-baselines3} \cite{rl-zoo3} \cite{liang2018rllib} \cite{hu2022marllib} \cite{bettini2024benchmarl} \cite{dekock2023mava} \cite{castro18dopamine} \cite{tianshou} \\

\bottomrule
\end{tabular}}
\end{table}


%% file: tables/component-responsibilities.tex
\begin{table*}[t]
\centering
\caption{Summary of Components and their Responsibilities}
\label{tab:ra-summary}
\renewcommand{\arraystretch}{0.9}
{\scriptsize
\begin{tabular}{@{}llp{11cm}@{}}
\toprule
\multicolumn{1}{c}{\textbf{Component}} & \multicolumn{1}{c}{\textbf{Container}} & \multicolumn{1}{c}{\textbf{Role}} \\ \midrule
\componentrefplain{sec:ra-experiment-orchestrator}{Experiment Orchestrator} & Framework &  Translates user-defined experiment specifications into the concrete execution process. \\
\componentrefplain{sec:ra-experiment-manager}{Experiment Manager} & Experiment Orchestrator & Sets up the Data Persistence and Monitor \& Visualization components and delegates the experiment execution to the Framework Orchestrator.\\
\componentrefplain{sec:ra-hyperparameter-tuner}{Hyperparameter Tuner} & Experiment Orchestrator & Generates candidate hyperparameter configurations. \\
\componentrefplain{sec:ra-benchmark-manager}{Benchmark Manager} & Experiment Orchestrator & Enables algorithm comparison under consistent experimental settings. \\ \midrule

\componentrefplain{sec:ra-framework-orchestrator}{Framework Orchestrator} & Framework Core & Initializes required framework components and coordinates their operation during training or evaluation. \\
\componentrefplain{sec:ra-lifecycle-manager}{Lifecycle Manager} & Framework Orchestrator & Controls the execution of the agent–environment interaction loop. \\
\componentrefplain{sec:ra-configuration-manager}{Configuration Manager} & Framework Orchestrator & Loads and validates the configurations. \\
\componentrefplain{sec:ra-multi-agent-coordinator}{Multi-Agent Coordinator} & Framework Orchestrator & Manages how multiple agents interact and learn. \\
\componentrefplain{sec:ra-distributed-execution-coordinator}{Distributed Execution Coordinator} & Framework Orchestrator & Allocates and deploys components across resources. \\ \midrule

\componentrefplain{sec:ra-agent}{Agent} & Framework Core & Implements the RL algorithm and learns through agent-environment interaction. \\
\componentrefplain{sec:ra-function-approximator}{Function Approximator} & Agent & Encodes the agent's
decision-making mechanism. \\
\componentrefplain{sec:ra-buffer}{Buffer} & Agent & Stores the collected experience.\\
\componentrefplain{sec:ra-learner}{Learner} & Agent & Updates the Function Approximator using
collected experience. \\ \midrule

\componentrefplain{sec:ra-environment}{Environment} & Framework Core & Encapsulates the simulation infrastructure with which the Agent interacts. \\
\componentrefplain{sec:ra-environment-core}{Environment Core} & Environment & Initializes the environment, applies agent actions, updates the environment state, and computes rewards. \\
\componentrefplain{sec:ra-environment-core}{Action Manager} & Environment Core & Processes and applies actions. \\
\componentrefplain{sec:ra-environment-core}{Observation Manager} & Environment Core & Collects observations from the simulator infrastructure. \\
\componentrefplain{sec:ra-environment-core}{Reward Manager} & Environment Core & Computes reward signals. \\
\componentrefplain{sec:ra-simulator}{Simulator} & Environment & Executes the probabilistic mechanism representing the real-world phenomenon, producing simulation traces.\\
\componentrefplain{sec:ra-simulator}{Simulator Adapter} & Environment & Connects the 
Environment Core with the underlying Simulator.\\ \midrule

\componentrefplain{sec:ra-data-persistence}{Data Persistence} & Utilities & Manages the storage and retrieval of the experiment data.\\
\componentrefplain{sec:ra-checkpoint-manager}{Checkpoint Manager} & Data Persistence & Saves and restores the experiment state. \\
\componentrefplain{sec:ra-environment-paramter-manager}{Env. Parameter Manager} & Data Persistence & Stores environment parameters that change over time.\\ \midrule

\componentrefplain{sec:ra-monitoring-and-visualization}{Monitoring \& Visualization} & Utilities & Tracks the learning process and generates result summaries. \\
\componentrefplain{sec:ra-renderer}{Renderer} & Monitoring\& Visualization & Generates visual frames of the environment state.\\ 
\componentrefplain{sec:ra-recorder}{Recorder} & Monitoring\& Visualization & Stores capture frames as videos or image sequences.\\
\componentrefplain{sec:ra-logger}{Logger} & Monitoring\& Visualization & Records raw experiment data.\\
\componentrefplain{sec:ra-reporter}{Reporter} & Monitoring\& Visualization & Transforms logged data into human-readable outputs.\\
\bottomrule
\end{tabular}}
\end{table*}

%% file: sections/reconstructing-patterns.tex
\section{Reconstructing RL Patterns}\label{sec:reconstructing-patterns}

To demonstrate the RA, we reconstruct typical RL patterns.

\subsection{Reconstructing Discrete Policy Gradient}
\begin{figure}[t]
    \centering
    \includegraphics[width=0.85\linewidth]{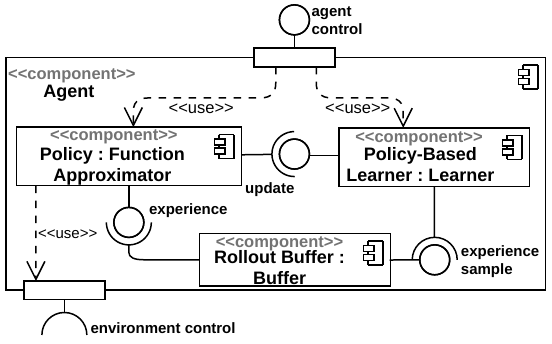}
    \vspace{-0.5em}
    \caption{Reconstruction of Discrete Policy Gradient}
    \label{fig:reconstruct-discrete-policy-gradient}
    \vspace{-0.75em}
\end{figure}

Discrete policy gradient RL methods directly learn a probability distribution over possible actions for any given state~\cite{sutton1999policy}. \figref{fig:reconstruct-discrete-policy-gradient} shows how such methods are instantiated from the RA.

In discrete policy gradient methods, the \componentref{sec:ra-function-approximator}{Function Approximator} is implemented by a stochastic, parameterized \component{Policy} over a finite action set. Given the current state of the \componentref{sec:ra-agent}{Agent}, the policy samples from a probability distribution to determine the next action, and the resulting experience is stored in a \componentref{sec:ra-buffer}{Rollout Buffer}, storing short trajectories that are consumed immediately and then discarded.
The \componentref{sec:ra-learner}{Policy-Based Learner} samples experience from the \componentref{sec:ra-buffer}{Buffer}, and updates the \component{Policy}.

\subsection{Reconstructing Q-learning}

\begin{figure}[t]
    \centering
    \includegraphics[width=0.85\linewidth]{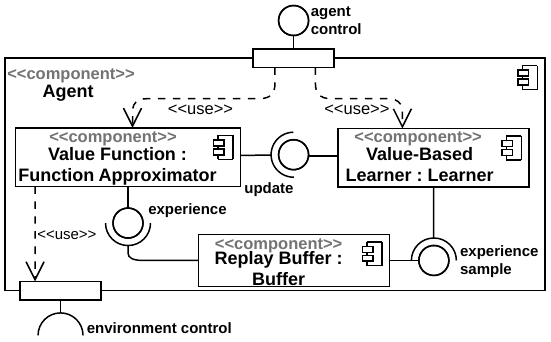}
    \vspace{-0.5em}
    \caption{Reconstruction of Q-learning}
    \label{fig:reconstruct-q-learning}
    \vspace{-0.75em}
\end{figure}

Q-learning is an example of value based methods, which---as opposed to policy-based ones---learn the value (the expected future reward) of states or actions to guide decisions~\cite{watkins1992q-learning}. \figref{fig:reconstruct-q-learning} shows how Q-learning is instantiated from the RA.

In Q-learning, the \componentref{sec:ra-function-approximator}{Function Approximator} is implemented by a \component{Value function} (here: Q-function~\cite{watkins1992q-learning}) and select actions based on the Q-values through an appropriate action selection strategy, e.g., $\epsilon$-greedy~\cite{watkins1989learning}. The resulting experience is stored in a \componentref{sec:ra-buffer}{Replay Buffer}, storing a large pool of transitions.
The \componentref{sec:ra-learner}{Value-Based Learner} samples mini-batches of experience from the \componentref{sec:ra-buffer}{Buffer}, computes temporal difference (TD) errors between current and target Q-values, and updates the \component{Policy} so that TD errors are minimized and by that, Q-value estimates improve.

\subsection{Reconstructing Actor-Critic}

\begin{figure}[t]
    \centering
    \includegraphics[width=0.92\linewidth]{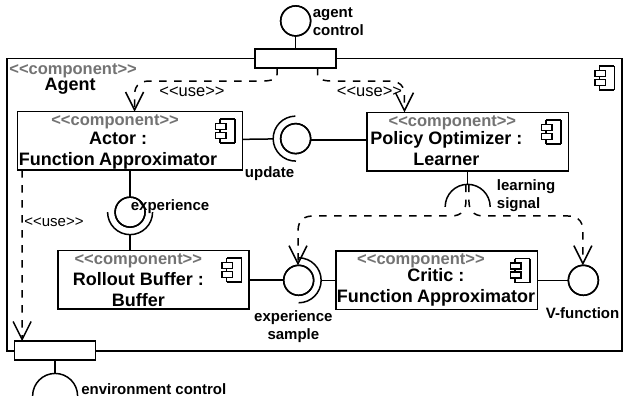}
    \vspace{-0.5em}
    \caption{Reconstruction of Advantage Actor-Critic (A2C)}
    \label{fig:reconstruct-a2c}
    \vspace{-0.75em}
\end{figure}

Actor-critic methods combine a policy-based \component{Actor} that prioritizes short-term learning, and a value-based \component{Actor} that prioritizes long-term learning.
\figref{fig:reconstruct-a2c} shows how advantage actor-critic (A2C), a specific flavor of actor-critic methods is instantiated from the RA.

In actor-critic methods, the \componentref{sec:ra-function-approximator}{Function Approximator} is instantiated twice: the \component{Actor} and the \component{Critic}.
The \component{Actor} implements policy-based learning. It conducts short rollouts and stores this experience in the \componentref{sec:ra-buffer}{Buffer}. The \component{Critic}, based on the \component{Actor}'s experience, estimates the state values, i.e., the V-function. Both the \component{Actor}'s experience and the \component{Critic}'s V-function is used by the \componentref{sec:ra-learner}{Policy Optimizer} to calculate the advantage of rollouts, and update the \component{Actor}'s policy by ascending the policy-gradient objective, which is a function of the advantage.

\subsection{Reconstructing Multi-Agent Learning}\label{sec:reconstruct-marl}

\begin{figure}[h]
    \centering
    \includegraphics[width=\linewidth]{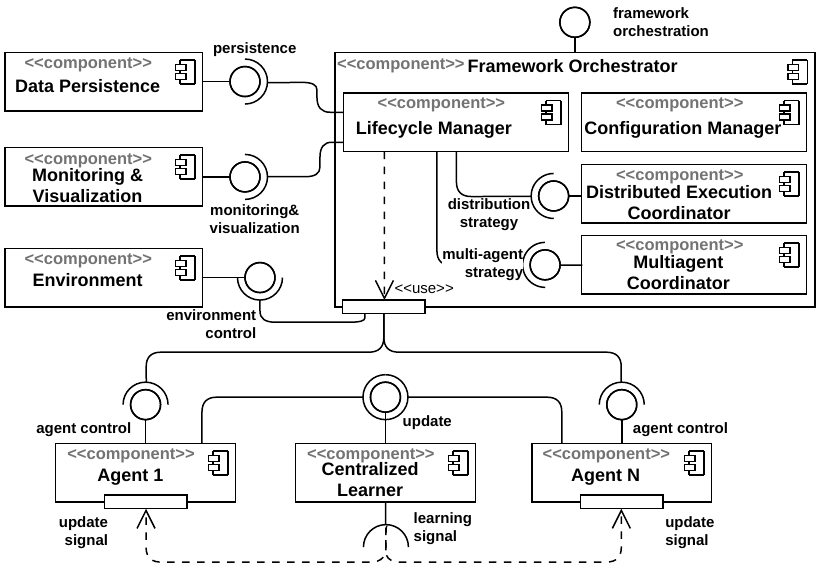}
    \vspace{-1.5em}
    \caption{Reconstruction of MARL (with centralized learning)}
    \label{fig:reconstruct-marl}
    \vspace{-0.5em}
\end{figure}

In decentralized multi-agent RL (MARL)~\cite{zhang2018fully}, agents learns independently on local information.
\figref{fig:reconstruct-marl} shows how the centralized flavor of MARL is instantiated from the RA.

In MARL, \componentref{sec:ra-agent}{Agent}s 1..N interact with the \componentref{sec:ra-environment}{Environment} individually, and produce update signals to either learn individually (decentralized MARL), or to be used by a \componentref{sec:ra-learner}{Centralized Learner} component (centralized MARL, shown in \figref{fig:reconstruct-marl}). \componentref{sec:ra-agent}{Agent}s are architected as shown in the previous examples, each containing their own \componentref{sec:ra-function-approximator}{Function Approximator} and \componentref{sec:ra-buffer}{Buffer}.
A \componentref{sec:ra-learner}{Centralized Learner} typically samples from all \componentref{sec:ra-buffer}{Buffer}s, computing joint learning signals (e.g., shared returns, team rewards), and updating all \componentref{sec:ra-agent}{Agent}'s \componentref{sec:ra-function-approximator}{Function Approximator}s.

The \componentref{sec:ra-multi-agent-coordinator}{Multi-Agent Coordinator} takes care of assembling joint actions for environment execution, distributing experience back to each agent, handling agent ordering, and maintaining policy-agent mappings for shared or individual policies.
In distributed MARL, the \componentref{sec:ra-distributed-execution-coordinator}{Distributed Execution Coordinator} takes care of deploying \componentref{sec:ra-agent}{Agent}s on different hardware components (e.g., different threads, GPUs, or clusters).

%% file: sections/evaluation.tex
\section{Results Evaluation and Quality Assessment}

Following best practices in GT~\cite{stol2016grounded,charmaz2006constructing}, we evaluate our results and the quality of our study by the following criteria.

\begin{description}[leftmargin=1em]
    \item[Credibility, i.e., \textit{is there \ul{sufficient} data to merit claims?}] In our study, we analyze 18 RL systems that support an array of RL techniques and use-cases. The resulting theory reached theoretical saturation relatively fast, after about 65\% of the corpus.
    Based on these observations, we conjecture that we analyzed a sufficient number of RL systems to render this study credible.
    
    \item[Originality, i.e., \textit{do the categories offer \ul{new} insights?}] Our study offers a comprehensive empirical view on how modern RL frameworks are architected. In addition, our study clarifies coarser-grained component groups (see \figref{fig:high-level} and the subsections of \secref{sec:ra}) to depart from informal colloquialisms of ``environments'' and ``frameworks.''
    
    \item[Resonance, i.e., \textit{does the theory make \ul{sense} to experts?}] To assess resonance, we gathered structured reflections from five experts (researchers, practitioners) knowledgeable in RL. We recruited the practitioners via convenience sampling and asked them the following questions. (1) \textit{When you read this reference architecture (RA), does anything stand out as inaccurate in your experience?} (2) \textit{Which parts of this RA match your daily reality, i.e., when working with RL?} (3) \textit{Are there components or relationships in this RA that do not fit your experience?} (4) \textit{Did anything in the RA help you see your experience with RL frameworks differently?}

    Experts unanimously responded that nothing stood out as inaccurate. Some experts mentioned components that do not fit their experience as they often implement custom components, but they could identify those components in the RA. About the benefits of the RA, one expert mentioned that they ``\textit{would use the RA to help me organize my code components that I have to implement myself}'' and that they ``\textit{would also use this RA when learning to work with a new framework, to map concepts onto the RA to help me understand how it works}.''
    Regarding modularity, one expert mentioned that ``\textit{the explicit identification of a Lifecycle Manager helped clarify the architectural role of the training loop},'' and that the RA highlighted how ``\textit{simulator integration can be treated as a reusable architectural pattern rather than a framework-specific detail}.'' Experts recognized opportunities the RA may bring, too, e.g., when ensuring convergence of RL algorithms: ``\textit{When convergence was difficult, it was often challenging to pinpoint whether the issue was architectural, environmental, or algorithmic. This reference architecture highlights how stronger modular separation could make debugging and experimentation clearer.}''
    
    We judge that the RA resonates well with practitioners as they articulate no misalignment with their practices and recognize value in the RA tied to their daily practices.
    
    It is important to note that this exercise is a mere resonance check rather than a thorough validation. As such, it does not allow for generalization. However, it indicates interpretive plausibility and experiential alignment.
    
    \item[Usefulness, i.e., \textit{does the theory offer \ul{useful} interpretations?}] Our study identifies architectural primitives of RL frameworks that are grounded in existing implementations, i.e., they allow for interpreting the RA in the practical context of RL. Beyond identifying components and component groups, our study allows researchers and practitioners to structure future RL systems and anticipate architectural trade-offs (e.g., highly modular design vs more integrated functionality).
\end{description}

\subsection*{Threats to validity}

\begin{description}[leftmargin=1em]
    \item[Internal validity] The inferred compositions of, and relationships between categories may reflect researchers' interpretations rather than inherent properties of the studied frameworks. We mitigated this threat through constant comparison and by actively searching for alternative explanations.

    \item[Construct validity] Researcher bias could threaten the validity of this study, e.g., architectural components may have been shaped by this effect. To mitigate this threat, we facilitated constant cross-verification steps and discussions among the researchers.
    Selection bias (e.g., missing important frameworks, and including less relevant ones) could threaten construct validity. To mitigate this threat, we relied on the best practices of grounded theory and extensively analyzed frameworks until we confirmed saturation.

    \item[External validity] GT is a non-statistical research genre and therefore, the results cannot be statistically generalized to a general population, e.g., to general machine learning frameworks and closed-source implementations. Since we reached saturation, we expect the RA to generalize to RL frameworks though.
\end{description}

%% file: sections/discussion.tex
\section{Discussion}\label{sec:discussion}

\subsection{The architectural tendencies of RL frameworks}

\begin{figure}[t]
    \centering
    \includegraphics[width=0.95\linewidth]{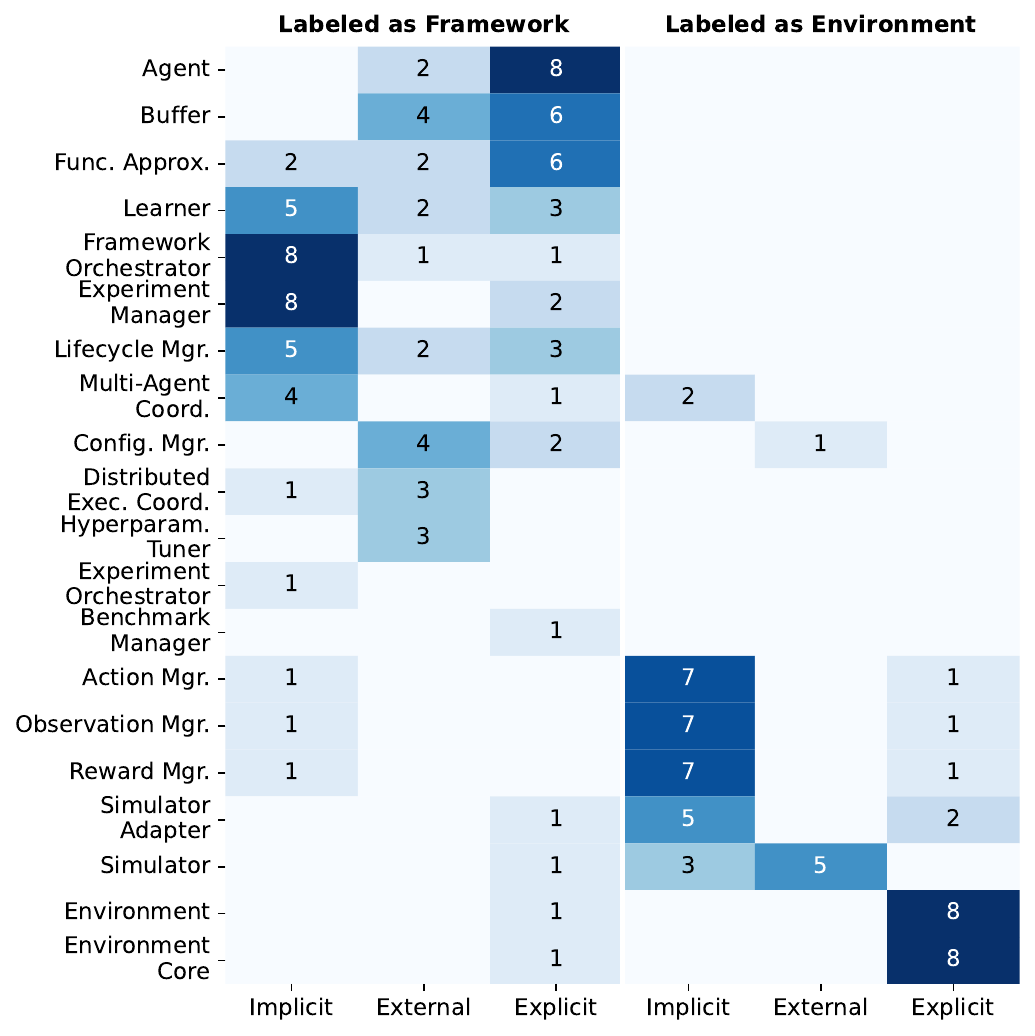}
    \caption{Implementations of RA components across the sampled RL systems. (\component{Utility} components not included.)}
    \label{fig:component-heatmap}
\end{figure}

The 18 frameworks and 28 RA components imply a total of 504 ($18\times28$) potential implementations across the sampled frameworks. In total, we find \xofy{252}{504}{} implemented components in the sampled frameworks and \xofy{252}{504}{} missing ones.
As shown in \figref{fig:component-heatmap}, RL systems labeled as ``frameworks'' and ``environments'' tend to implement complementary functionality. (The figure does not include \componentref{sec:ra-utilities}{Utilities}, which shows a more homogeneous coverage of components across framework- and environment-type RL systems.)
Framework-type RL systems (e.g., Acme~\cite{hoffman2020acme} and RLlib~\cite{liang2018rllib}) tend to implement the \componentref{sec:ra-agent}{Agent} components (e.g., \componentref{sec:ra-buffer}{Buffer} and \componentref{sec:ra-function-approximator}{Function Approximator} -- \figref{fig:ra-agent}) and \componentref{sec:ra-framework-orchestrator}{Framework Orchestrator} components (e.g., \componentref{sec:ra-lifecycle-manager}{Lifecycle Manager} and \componentref{sec:ra-multi-agent-coordinator}{Multi-Agent Coordinator} -- \figref{fig:ra-framework-orchestrator}); and exhibit \xofy{75}{90}{} coverage on such components.\footnote{Implementation details are available in the data package.}
Environment-type RL systems (e.g., Gymnasium~\cite{towers2025gymnasium} and PettingZoo~\cite{terry2021pettingzoo}) tend to be restricted to implement the \componentref{sec:ra-environment}{Environment} group (\figref{fig:ra-environment}) and exhibit \xofy{55}{56}{} coverage on such components.

These figures highlight the \textbf{complementary architectural tendencies} of RL systems colloquially labeled as ``environments'' and ``frameworks,'' and hint at the \textbf{importance of considering both types} when designing RL-based software.

\begin{recommendationframe}
    We recommend developers of RL-enabled software to consider the complementary feature set of RL systems that implement environments and those that implement framework components. Complex software solutions may necessitate integrating both types.
\end{recommendationframe}

\addblockend{}

\input{tables/ADD-localization}

\subsection{The role of external libraries}

We find \xofy{123}{252}{} explicitly implemented components (i.e., the responsibility of an RA component is clearly assigned to an existing component); \xofy{47}{252}{} that are implemented through an external library (i.e., explicit isolation of the responsibility but the implementation is deferred); and \xofy{82}{252}{} implicitly implemented ones (i.e., responsibilities are lumped into other components).

\figref{fig:component-heatmap} shows that external implementations can be found in the vast majority of RA component types, especially the ones that are \componentref{sec:ra-framework}{Framework}-related. This tendency hints at the existence of external libraries that can act as viable building blocks of RL software.
For example, the TensorBoard~\cite{tensorboard-website} is a recurring library to implement the \componentref{sec:ra-reporter}{Reporter} component, e.g., in Unity ML-Agents~\cite{juliani2020}, Isaac Gym~\cite{makoviychuk2021isaac}, and Dopamine~\cite{castro18dopamine}. Hydra~\cite{yadan2019hydra} is often used to implement the \componentref{sec:ra-configuration-manager}{Configuration Manager} component, e.g., in Isaac Lab~\cite{mittal2025isaac}, BenchMARL~\cite{bettini2024benchmarl} and Mava~\cite{dekock2023mava}.
This finding is in line with recent reports on the fundamental enabler role of open-source software stacks in ML~\cite{tan2022exploratory} and the fast-moving landscape of libraries in machine learning (ML)~\cite{nguyen2019machine}. In such a landscape, \textbf{clear architectural guidelines, such as RAs are in high demand}. For example, a recent study by \textcite{lariosvargas2020selecting} shows that the impact of a wrongly chosen external library depends on the library's role in the software architecture, and that experts tend to address this problem by evaluating the architectural alignment of external libraries in the prototype phase of software.

In our sample, \xofy{18}{47}{} external libraries originate directly from the same development team or ecosystem.
For example, in RLlib~\cite{liang2018rllib}, multiple components are realized through libraries from the same team, Anyscale, e.g., Ray Tune~\cite{liaw2018tune} to implement the \componentref{sec:ra-hyperparameter-tuner}{Hyperparameter Tuner} and Ray Core~\cite{moritz2018ray} to implement the \componentref{sec:ra-distributed-execution-coordinator}{Distributed Execution Coordinator}. Similarly, Acme~\cite{hoffman2020acme} uses libraries from Google DeepMind, e.g., Launchpad~\cite{yang2021launchpad} to implement the \componentref{sec:ra-distributed-execution-coordinator}{Distributed Execution Coordinator} and Reverb~\cite{cassirer2021reverb} to implement \componentref{sec:ra-buffer}{Buffer}s.

\begin{recommendationframe}
We recommend that adopters assess the supporting ecosystem as reliable external libraries may improve the quality of the developed RL-enabled software.
\end{recommendationframe}

\subsection{Localizing Architectural Design Decisions (ADD) in RL}

The RA in this work allows for localizing the Architectural Design Decisions (ADD) for RL by \textcite{ntentos2024supporting}. ADDs represent the design choices that influence how an RL framework is constructed and operated. \textbf{By understanding which components are affected by a specific ADD, architects can evaluate the feasibility and impact of a decision more precisely.}
For example, components relying on external dependencies may not allow for the same liberty as custom implementations (e.g., versioning may be more rigorous for the sake of backward compatibility).
\tabref{tab:discussion-add} maps ADDs onto the RA components that they primarily affect.

The \textit{Model Architecture} ADD determines the number of \componentref{sec:ra-agent}{Agent}s, their internal structure, and whether multi-agent coordination (by the \componentref{sec:ra-multi-agent-coordinator}{Multi-Agent Coordinator}) is required. 
For example, A monolithic model uses a single \componentref{sec:ra-agent}{Agent}, and a multi-agent architecture requires multiple \componentref{sec:ra-agent}{Agent}s and a \componentref{sec:ra-multi-agent-coordinator}{Multi-agent Coordinator} to manage agent interactions.
The \textit{Model Training} ADD determines where and how learning updates are performed. For example, centralized training with decentralized execution MARL uses a single \componentref{sec:ra-learner}{Centralized Learner} accessing multiple \componentref{sec:ra-buffer}{Buffer}s, and the \componentref{sec:ra-multi-agent-coordinator}{Multi-agent Coordinator} handles joint action assembly and experience distribution across agents. 
Distributed training requires the \componentref{sec:ra-distributed-execution-coordinator}{Distributed Execution Coordinator} to deploy components across distributed resources.
The \textit{Checkpoints} ADD determines whether, what, and when to save the experiment states. The \componentref{sec:ra-checkpoint-manager}{Checkpoint Manager} handles this decision by deciding when (e.g., every N steps) and what (e.g., training states, model parameters) to save.
The \textit{Transfer Learning} ADD determines whether to train from pre-trained models or from scratch. If \textit{Transfer Learning} is enabled, the \componentref{sec:ra-checkpoint-manager}{Checkpoint Manager} loads the pre-trained model parameters into the \componentref{sec:ra-agent}{Agent}.
The \textit{Distribution Strategy} ADD determines where and how to use distributed training.
The \componentref{sec:ra-distributed-execution-coordinator}{Distributed Execution Coordinator} handles this decision by allocating and deploying components across resources.
The \textit{Hyperparameter Tuning} ADD determines whether to use hyperparameter tuning or not. 
If \textit{Hyperparameter Tuning} is enabled, the \componentref{sec:ra-hyperparameter-tuner}{Hyperparameter Tuner} automates the search over hyperparameter configurations to find the optimal ones.

\begin{recommendationframe}
We recommend RL architects use our RA to localize design decisions for better assessment and evaluation of the implications of their decisions.
\end{recommendationframe}

%% file: tables/ADD-localization.tex
\begin{table*}
\centering
\caption{Localizing Architectural Design Decisions}
\vspace{-0.5em}
\label{tab:discussion-add}
\renewcommand{\arraystretch}{0.8}
{\footnotesize
\begin{tabular}{@{}p{2.7cm}p{6.3cm}p{8.3cm}@{}}
\toprule
\multicolumn{1}{c}{\textbf{ADD}} & \multicolumn{1}{c}{\textbf{Definition}} & \multicolumn{1}{c}{\textbf{Affected RA components}} \\ \midrule

Model Architecture & The structural organization of RL models & \componentrefplain{sec:ra-agent}{Agent}, \componentrefplain{sec:ra-multi-agent-coordinator}{Multi-Agent Coordinator} \\

Model Training & The organization of learning across agents & \componentrefplain{sec:ra-multi-agent-coordinator}{Multi-Agent Coord.}, \componentrefplain{sec:ra-distributed-execution-coordinator}{Distributed Execution Coord.}, \componentrefplain{sec:ra-learner}{Learner}, \componentrefplain{sec:ra-buffer}{Buffer} \\

Checkpoints & Whether RL models' states are saved during learning & \componentrefplain{sec:ra-checkpoint-manager}{Checkpoint Manager}  \\

Transfer Learning & Whether to use pre-trained models or train from scratch & \componentrefplain{sec:ra-checkpoint-manager}{Checkpoint Manager}, \componentrefplain{sec:ra-agent}{Agent} \\

Distribution Strategy & Whether and how to use distributed training & \componentrefplain{sec:ra-distributed-execution-coordinator}{Distributed Execution Coordinator} \\

Hyperparameter Tuning & Whether to use hyperparameter tuning or not & \componentrefplain{sec:ra-hyperparameter-tuner}{Hyperparameter Tuner} \\
\bottomrule
\end{tabular}}
\vspace{-0.5em}
\end{table*}

%% file: sections/conclusion.tex
\section{Conclusion}\label{sec:conclusion}

In this paper, we propose a reference architecture of RL frameworks based on our empirical investigation of 18 widely used open-source implementations. Our work allows for better-informed design decisions in the development and maintenance of RL frameworks and during the integration of RL frameworks into software systems.
We plan to maintain our reference architecture by continuously analyzing new RL frameworks.
Future work will focus on a more detailed evaluation of the RA to assess its understandability, effectiveness in solving practical RL problems, and usefulness to practitioners.
In addition, we plan to provide a reference implementation that RL system developers can use as reasonable abstractions and starting points for their implementations.

%% file: sections/ack.tex
\section*{Acknowledgment}

We truly appreciate the insightful remarks of the three anonymous reviewers who helped us improve the original submission; and our colleagues who provided structured feedback in the resonance check phase of this work.